\documentclass[12pt,a4paper]{article}
\usepackage[preprint]{dfttrob}
\usepackage[numbers,sort&compress]{natbib}
\usepackage{latexuseful2e}
\usepackage{amsmath}
\usepackage{graphicx}

\dfttnum{DFTT 02/2003\\17 January 2003}

\begin{document}

\title{Open problems in forward-backward multiplicity correlations in
	hadron-hadron collisions in the TeV region}
\author{A. Giovannini and R. Ugoccioni\\
 \it Dipartimento di Fisica Teorica and INFN - Sezione di Torino\\
 \it Via P. Giuria 1, 10125 Torino, Italy}
\maketitle

\begin{abstract}
Continuing previous work on forward-backward multiplicity correlation
properties in proton-proton collisions  in the framework  of the
weighted superposition model  of two components (each one described by
a negative binomial multiplicity distribution) 
with the addition of the leakage parameter which controls clan spreading
from one hemisphere to the
opposite one, we examine E735 data on the c.m.\ energy dependence 
of the total correlation strength and of the forward variance at fixed 
total multiplicity. A comparison with the Chou-Yang approach to the
problem is presented and extrapolations of the mentioned variables 
at LHC c.m.\ energy  in possible scenarios in the new energy domain
are discussed.
\end{abstract}

\newcommand{\btotal}{b_{\text{total}}}
\newcommand{\psoft}{p_{\text{soft}}}
\newcommand{\psemi}{p_{\text{semi-hard}}}
\newcommand{\zn}{\avg{z^2}_n}


As discussed in Ref.~\cite{RU:FB}, the c.m.\ energy dependence of the
forward-backward multiplicity correlation (FBMC) strength, $\btotal$,
in \ee\ annihilation and \pp\ collisions can be understood as the result
of the superposition of two components or classes of events.
The two components are, respectively, 2- and 3-jet event classes
in \ee\ annihilation, soft (without mini-jets) and semi-hard (with
mini-jets) classes in \pp\ collisions.
A general formula was given for $\btotal$ which does not depend on the
specific form of the multiplicity distribution (MD) in each component,
but only on the first two moments, the FBMC strength in each
component, $b_1$ and $b_2$, and the weight factor, $\alpha$.

In the case of \ee\ annihilation, the correct value of $\btotal$ at
LEP energy was reproduced under the experimental conditions
$b_{\text{2-jet}} \approx b_{\text{3-jet}} \approx 0$.
This fact was considered as a successful test of the superposition
mechanism itself.

The situation was found to be quite different in \pp\ collisions. Here
it was shown that the correct energy dependence of $\btotal$ can be
determined by assuming, in addition to the above mentioned parameters,
the explicit form of the MD in the soft and semi-hard components
(i.e., two negative binomial (NB) MD's with different parameters) and
related clan structure, and by introducing the corresponding particle
leakage parameters $\psoft$ and $\psemi$, which control clan spreading
over both hemispheres. The leakage parameter is indeed defined as the
fraction of particles within one clan which remain in the same
hemisphere where the clan was produced.

Within this framework, the energy dependence of $\btotal$ was then
extrapolated in the TeV region by examining three different scenarios
(see \cite{combo:prd} for details)
characterised by the same soft component structure satisfying KNO
scaling, and by different semi-hard component behaviours: 1)
obeying again KNO scaling, or 2) with strong KNO scaling violation or 3)
with a QCD inspired behaviour.
A clear bending of $\btotal$ was visible in all examined scenarios. It
was remarked that an early saturation of $\btotal$ toward 1 in the
semi-hard component would require a fast increase with energy of
particle leakage, i.e, a decrease of the corresponding leakage
parameter $\psemi$, and would favour strong KNO scaling violation. 

The aim of this paper is to discuss in the mentioned framework the
E735 Collaboration results \cite{FB:E735} on c.m.\ energy dependence
of the FBMC strength, $\btotal$, and of the forward variance at fixed
total multiplicity $n$, $d^2_{n_F}(n)$, obtained at Tevatron.
It should be pointed out that
\begin{equation}
	\btotal \equiv \frac{ \avg{ (n_F-\bar n_F)(n_B-\bar n_B) }}{ 
	    \sqrt{ \avg{(n_F-\bar n_F)^2}\avg{(n_B-\bar n_B)^2} }}
		= \frac{ D^2_n - 4\avg{d^2_{n_F}(n)} }{ D^2_n +
	    4\avg{d^2_{n_F}(n)} } ,
		                                               \label{eq:1}
\end{equation}
where $\avg{~}$ indicates an average over all events, and $D_n$ is the
dispersion of the MD. Furthermore, let us introduce the variable
\begin{equation}
	\zn \equiv \avg{ n_F-n_B }_n = 4d^2_{n_F}(n) ,
                                                   \label{eq:2}
\end{equation}
where $\avg{~}_n$ indicates the average over all events at fixed $n$.
It is clear that variable (\ref{eq:2}) works at a deeper level of
investigation, being variable (\ref{eq:1}) related to the average of
(\ref{eq:2})  over all multiplicities. Variable (\ref{eq:1}) is
particularly interesting for global properties of the collisions
related to average $n$.

Variable (\ref{eq:1}), in the weighted two-component
superposition model summarised above, can be expressed as follows:
\begin{equation}
	\btotal = 	\frac{\alpha \frac{b_1}{(1+b_1)} {D^2_{n,1}} +
				(1-\alpha) \frac{b_2}{(1+b_2)} {D^2_{n,2}} +
					\frac{1}{2}\alpha(1-\alpha)(\nbar_{2} - \nbar_{1})^2}
			{\alpha  \frac{1}{(1+b_1)}{D^2_{n,1}} +
				(1-\alpha)  \frac{1}{(1+b_2)}{D^2_{n,2}} +
					\frac{1}{2}\alpha(1-\alpha)(\nbar_{2} - \nbar_{1})^2} ,
                                             \label{eq:3}
\end{equation}
where the single component FBMC strength is
\begin{equation}
	b_i = \frac{ 2\nbar_i p_i(1-p_i) }{
		            \nbar_i + k_i - 2 \nbar_i p_i(1-p_i) }, \label{eq:4}
\end{equation}
with $i=\text{soft},\text{semi-hard}$; $k$ is the parameter of the
NBMD which is related to the dispersion by
$k^{-1}=(D^2_n - \nbar)\nbar^{-2}$.

Variable (\ref{eq:2}) in turn can be written as
\begin{equation}
	\zn = 4 d^2_{n_F,1}(n) \frac{\alpha P_1(n)}{P(n)} +
	     4 d^2_{n_F,2}(n) \frac{(1-\alpha) P_2(n)}{P(n)} ,
\end{equation}
where
\begin{equation}
	P(n) = \alpha P_1(n) + (1-\alpha) P_2(n)
\end{equation}
is the total MD, with $P_1(n)$ and $P_2(n)$ the two component MD's,
respectively.

\subsection*{energy dependence of $\btotal$}

\begin{figure}[h]
  \begin{center}
  \mbox{\includegraphics[width=\textwidth]{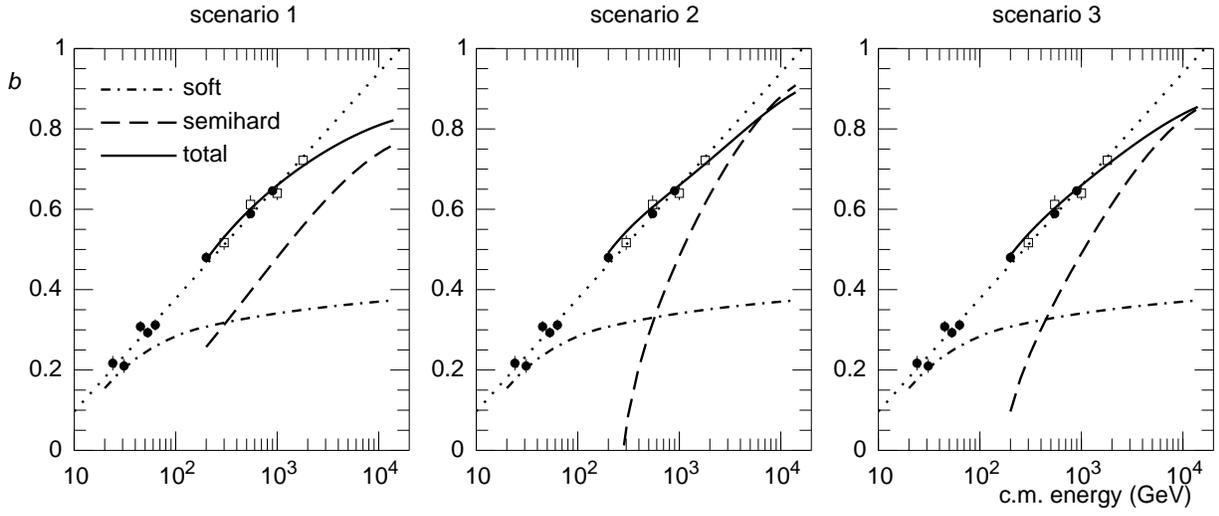}}
  \end{center}
  \caption{Energy dependence of the correlation coefficients 
        for each component (soft and semi-hard) 
        and for the total distribution in $p\bar p$ collisions.
				The dotted line is a fit to 
        experimental values \cite{UA5:correlations,FB:E735}.
	}\label{fig:1}
  \end{figure}

It is  found that the  points at 1000 GeV and 1800 GeV  from E735 
Collaboration  have the same energy dependence 
(they lie on the same straight line) as the other data in  
the GeV region \cite{FB:E735}.
In order to include the point at
1800 GeV the three extrapolated  scenarios for the semi-hard component 
discussed in Ref.~\cite{RU:FB}  are reexamined. 
Results are shown in Fig.~\ref{fig:1}.

In general, one can conclude that
the leakage parameter for the semi-hard
component, $\psemi$, must decrease, and accordingly particle leakage 
increase, in all scenarios. 
To be quantitative, we have found that satisfactory results are
obtained by taking the
leakage parameter for the soft component
energy independent and equal to 0.8, 
as argued in \cite{RU:FB}, and taking tentatively
$\psemi = 0.84 - 0.07 \log (\sqrt{s} / 200)$ for
$\sqrt{s} > 200~\text{GeV}$;
keeping this energy dependence for $\psemi$ at all energies, the
curves in Fig.~\ref{fig:1} have been extrapolated to 14 TeV.
It should be noticed that in scenario 2, characterised by a semi-hard
component with strong KNO scaling violation, the FBMC strength becomes
less steep with the increase of the c.m.\ energy and its saturation
toward 1 (as that of $\btotal$) quicker than in the other two
scenarios.

In conclusion, a linear behaviour of $\btotal$
with c.m.\ energy is incompatible with our approach above 2.5 TeV in
scenario 1, above 3.5 TeV in scenario 2 and above 5 TeV in scenario 3,
i.e., the leakage parameter energy dependence cannot be adjusted to
such situation in the various scenarios without spoiling the model
itself. On the contrary, if such a linear behaviour were found
experimentally at LHC, it could be indicative of the onset of a third
component (class of events).

\subsection*{$\zn$ vs $n$ dependence}

\begin{figure}[h]
  \begin{center}
  \mbox{\includegraphics{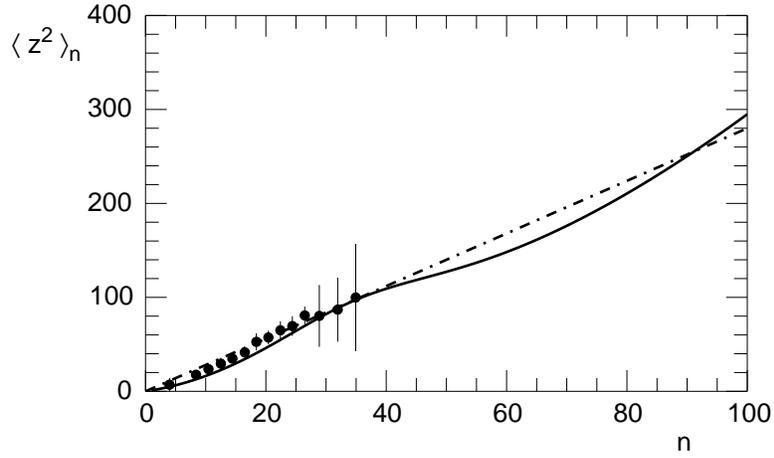}}
  \end{center}
  \caption{$\zn$ vs $n$ at 900 GeV in the interval $1<|\eta|<4$.
		    Data points are from UA5 Collaboration
		    \cite{UA5:correlations},
				the solid line is the result of our model in $0<|\eta|<4$,
				the dash-dotted line is a linear fit.
	}\label{fig:2}
  \end{figure}

Figure \ref{fig:2} shows experimental data from UA5 Collaboration 
\cite{UA5:correlations} at
900 GeV in $1<|\eta|<4$ together with the result of calculations in
the  present approach (solid line); the dashed line is a linear fit
according to the cluster model of Chou and Yang \cite{Chou:FB},
already discussed in \cite{RU:FB}.
Below $n\approx 40$, where data are available, it is quite hard to
distinguish the two model predictions; at $n\approx 40$ our approach
shows a ``hump''.
In view of the lack of sufficiently precise data in this domain, no
conclusions can be drawn.

\begin{figure}
  \begin{center}
  \mbox{\includegraphics{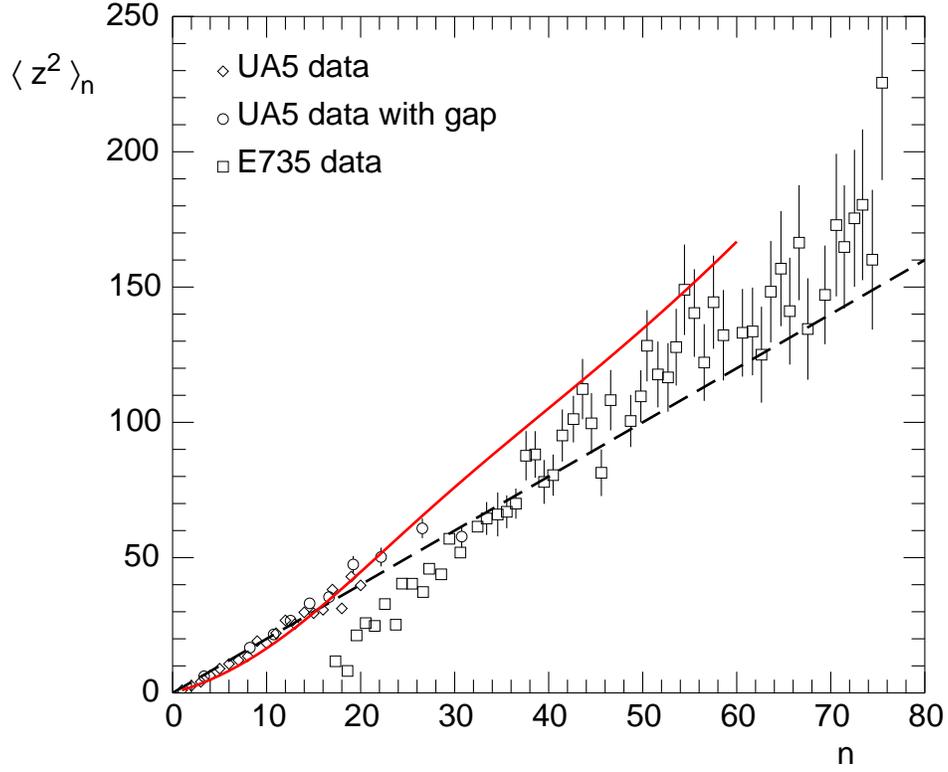}}
  \end{center}
  \caption{$\zn$ vs $n$ at 546 GeV. Data from
		      \cite{UA5:correlations,Chou:FB} and \cite{FB:E735} are compared
					with each other, with the prediction of our model (solid
		      line) and with a linear fit (dashed line).
	}\label{fig:3}
  \end{figure}

It should be pointed out that at higher c.m.\ energy (1.8 TeV), the
E735 data also show at $n\approx 40$ a qualitative picture like 
a hump. However, a quantitative comparison is problematic for two
reasons:
a) our calculations are based on extrapolations in full phase-space
while data refer to the interval $|\eta|<3.25$, for which no MD has
been published (on the contrary, available MD data from UA5
allowed us to compare our model's predictions at Sp\=pS);
b) at lower c.m.\ energy (546 GeV, see Fig.~\ref{fig:3}) we noticed a
discrepancy between UA5 results \cite{UA5:rep} and E735 results; 
for completeness, in
Fig.~\ref{fig:3} our results and a linear fit are also shown.

\begin{figure}
  \begin{center}
  \mbox{\includegraphics{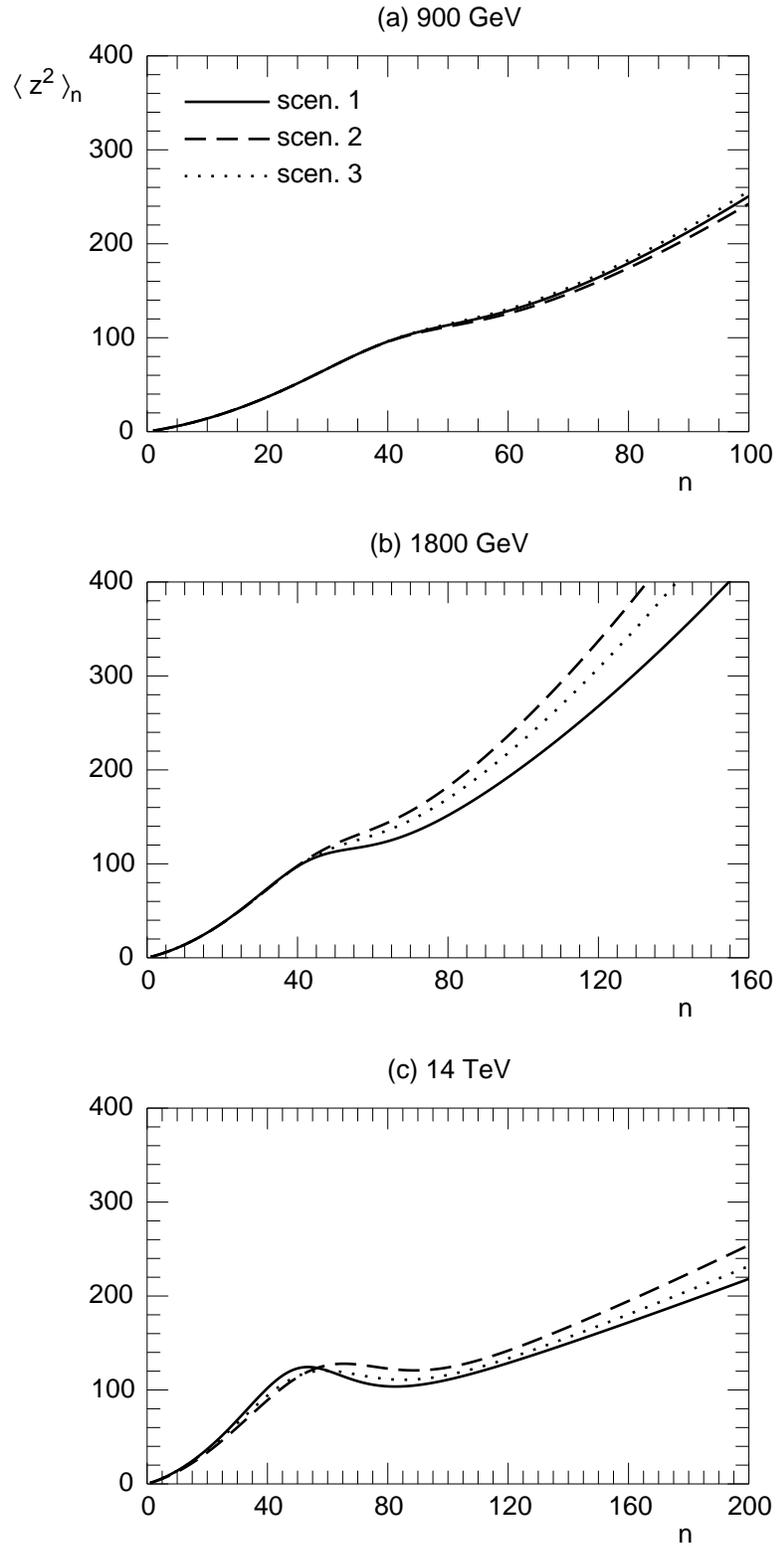}}
  \end{center}
  \caption{Results for $\zn$ vs $n$ in full phase-space
           for different c.m.\
	         energies  in different scenarios.}\label{fig:4}
  \end{figure}

The three scenarios we have discussed previously  do not show  any remarkable 
difference in the GeV region as far as the $\zn$ vs $n$ dependence is
concerned (see Fig~\ref{fig:4}a, where $\zn$
vs $n$ is plotted in the three
scenarios at 900 GeV in full phase-space).
In Fig.~\ref{fig:4}b the same plot is shown at 1800 GeV.
Differences in  the three 
scenario predictions  become more evident for $n$ larger than 40 and the hump
is more visible.
In Fig.~\ref{fig:4}c  at 14 TeV the hump becomes even more 
visible and in addition its maximum varies with the scenario.

In conclusion, the behaviour of $\zn$, i.e., the two
different sides of the hump appearing in the plot of $\zn$ vs
$n$, which is remarkable at 1800 GeV, confirms in our view the
presence of two components (samples of two classes of events) and
the importance of the role of the semi-hard component at this energy.
The question remains whether this is also the indication of 
the occurrence of a phase transition \cite{Gutay:plasma}.

\section*{Acknowledgements}
One of us (R.U.) would like to thank L. Gutay for discussion
on E735 results.

\bibliographystyle{prstyR}  
\bibliography{abbrevs,bibliography}

\newpage

\end{document}